\documentclass [{onecollarge,other options}] {svjour2}
\usepackage{graphicx}
\usepackage{dcolumn}
\usepackage{bm}
\usepackage{hyperref}
\usepackage{amssymb}
\usepackage{hyperref}
\RequirePackage[normalem]{ulem} 
\RequirePackage{color}\definecolor{RED}{rgb}{1,0,0}\definecolor{BLUE}{rgb}{0,0,1} 

\begin{document}
\newcommand{\be}{\begin{equation}}
\newcommand{\ee}{\end{equation}}
\newcommand{\ba}{\begin{eqnarray}}
\newcommand{\ea}{\end{eqnarray}}
\newcommand{\Gam}{\Gamma[\varphi]}
\newcommand{\Gamm}{\Gamma[\varphi,\Theta]}
\thispagestyle{empty}

\title{Entanglement entropies in the ground states of
helium-like atoms}

\author{Przemys\l aw Ko\'scik, Anna Okopi\'nska
Institute of Physics,  Jan Kochanowski University\\
ul. \'Swi\c{e}tokrzyska 15, 25-406 Kielce, Poland}

\maketitle

 \begin{abstract}
We examine the  entanglement in  the ground states of helium and
helium-like ions using an original Hylleraas expansion. The von
Neumann and linear entropies of the reduced density matrix are
accurately computed by performing the Schmidt decomposition of the
 S singlet spatial wavefunctions.  The results presented
are more accurate than currently available in published literature.

\end{abstract}

\section{Introduction}


In recent years there has been a lot of interest in entanglement
properies of few-body systems. Besides an important role in quantum
information technology, entanglement also attracts attention in view
of the problem of quantifying the amount of correlations in the
systems. Mainly, two-particle model systems confined by various
potential have been studied in this respect
\cite{2,3,3.1,4.1,4,5,7,8,9,9.1,10,11}. The von Neumann entropy (vN)
of the one-particle reduced-density matrix (RDM) is accepted as a
reliable entanglement measure for systems of two indistinguishable
particles \cite{vn}. At the same time, the vN entropy determines the
strength of correlation in the system. The linear entropy, being the
lower approximation of the vN entropy, is also used to this end,
since it can be more easily calculated without knowing the spectrum
of the RDM.

The quantum information content of two-electron atoms become also
interesting for chemists. For example, Manzano \textsl{et al.}
\cite{4}, Dehesa \textsl{et al.} \cite{he1} and Benenti \textsl{et
al.} \cite{he3} have studied the entanglement properties of the
ground and excited states of the helium atom. Y.C Lin \textsl{et
al.} \cite{he2} explored entanglement in the ground and excited
states of the helium atom and helium-like ions
, using configuration interaction wave functions  constructed with
B-spline basis. Very recently, C. H Lin \textsl{et al.} \cite{he4}
studied entanglement in the ground states of helium and the hydrogen
negative ion, establishing the values of linear entropies with
relatively small uncertainties. In most of the studies mentioned
above, only the linear entropy was used to quantify the amount of
the entanglement. We are aware of only four papers \cite{he3},
\cite{he5}, \cite{huang}, \cite{Hofer}, where the results for the
von Neumann  entropy were reported. However, there are large
discrepancies between the results of those papers.
This was our motivation for performing accurate calculations of entropies for the helium isoelectronic series. We calculated the vN and linear entropy
, basing on the Schmidt decomposition of the two-particle wavefunction. In our calculations, we used  an original correlated Hylleraas basis which allows determination of the wavefunction within a reasonable accuracy with relatively low computational cost.

This paper is organized as follows. In section \ref{12} we briefly
discuss the procedure to analyze the entanglement properties of the singlet
S-symmetry states. Section \ref{res} outlines our results, and some
concluding remarks are placed in Section \ref{summ}

\section{Methods}\label{12}

The Hamiltonian of atomic systems with two electrons and a nucleus of charge $Z e$ is given by
\begin{eqnarray} H=-{1\over 2}
\bigtriangledown_{1}^{2}-{1\over 2} \bigtriangledown_{2}^{2}-{Z\over
r_{1}}-{Z\over r_{2}}+{1\over r_{12}},\label{hamff}\end{eqnarray}
where  atomic units are used.
As mentioned before, we are interested in the singlet ground-state, the
spatial wave-function of which depends only on the radial coordinates $r_{1}, r_{2}$ and the  inter-electronic angle $\theta$. The Schmidt decomposition of the wave-function has a form
 \cite{sh}
\begin{eqnarray}\psi(\textbf{r}_{1},\textbf{r}_{2})\equiv\psi(r_{1},r_{2},\cos\theta)=\sum_{{n,l=0}}^{\infty}\sum_{m=-l}^{m=l}a_{nl} u_{nlm}^{*} (\textbf{r}_{1})u_{nlm}
(\textbf{r}_{2}),\label{Schmid}
\end{eqnarray}
 with $ u_{nlm}(\textbf{r})= {v_{nl}
(r)Y_{lm}(\theta,\varphi)\over r} $ and $a_{nl}={4\pi k_{nl}\over
2l+1}$, where $Y_{lm}$  are the spherical harmonics and $l$ and $n$
are the angular and principal quantum numbers, respectively. Both
the radial orbitals $v_{nl} (r)$ and the coefficients $k_{nl}$ are
real and can be determined by the following integral equations
\cite{Wang} \be \int_{0}^{\infty} f_{l}(r_{1},r_{2}) v_{nl} (r_{2})
dr_{2}=k_{nl} v_{nl}(r_{1})\label{integr}\ee with
 \begin{eqnarray}
f_{l}(r_{1},r_{2})=r_{1} r_{2}{2l+1\over 2}
\int_{0}^{\pi}\psi(r_{1},r_{2},\cos\theta)P_{l}(\cos\theta)\sin\theta
d\theta,\label{poloi1}
\end{eqnarray}
where $P_{l}$ are the Legendre polynomials. The natural orbitals
$u_{nlm}(\textbf{r})$  are the eigenvectors of the spatial RDM
$$\rho(\textbf{r},\textbf{r}^{'})=\int
[\psi(\textbf{r},\textbf{r}_{1})]^{*}\psi(\textbf{r}^{'},\textbf{r}_{1})d\textbf{r}_{1},$$
the eigenvalues of which, $\lambda_{nl}$, are related to the
expansion coefficients in (\ref{Schmid}) by $\lambda_{nl}=a_{nl}^2$. The natural orbitals $\{u_{nlm}(\textbf{r})\}_{m=-l}^{m=l}$ correspond to the same occupation number $\lambda_{nl}$,
which means that  $2l+1$-fold degeneracy occurs and, therefore,  the
normalization condition gives $\sum_{nl}(2l+1)\lambda_{nl}=1$. Using
Eq. (\ref{Schmid}), the identity
$[Y_{l,m}(\theta,\varphi)]^{*}=(-1)^m Y_{l,-m}(\theta,\varphi)$ and
the spin singlet function representation $\chi_{S}={\frac{1}{\sqrt{2}}}(\alpha(1)\beta(2)-\alpha(2)\beta(1))$,
where $\alpha(\beta)$ denotes the up (down) 
spin, the Slater decomposition of the total two-electron singlet
S-state wavefunction can be easily inferred, namely:
\begin{eqnarray}\Psi(\xi_{1},\xi_{2})=\sum_{{n=0\atop
l=0}}^{\infty}a_{nl}SD[ u_{nl0} \alpha, u_{nl0}\beta] +
\sum_{i,j=\{\alpha,\beta\}\atop i\neq j}\sum_{{n=0\atop
l=1}}^{\infty}\sum_{m=1}^{l}\nu a_{nl} SD[u_{nlm}^{*} i, u_{nlm}
j]\label{1},\end{eqnarray} where $\nu=1(-1)$ for $i=\alpha (\beta)$,
and $SD$ denotes a Slater determinant made out of two spin orbitals,
$SD[\phi i,\varphi j]=2^{-{1\over 2}}\left|\begin{array}{cc}\phi
(\textbf{r}_{1})i(1)&\varphi (\textbf{r}_{1})j(1)\\\phi
(\textbf{r}_{2})i(2)&\varphi
(\textbf{r}_{2})j(2)\end{array}\right|$.

Entanglement in pure states is usually quantified by the vN entropy
of the RDM, which in the case of the singlet sates takes the form
 \be \textbf{S}=-\mbox{Tr}[\rho\log_{2}\rho],\label{polo}\ee or the 
 linear entropy
 \be \textbf{L}=1-\mbox{Tr}[\rho^{2}],\label{polo1}\ee
 which both vanish when the corresponding total
two-electron wavefunction can be expressed as a single determinant
\cite{pla}. The linear entropy can be calculated without determining
the occupation numbers, as the spatial purity $\mbox{Tr}[\rho^2]$
can be expressed by the twelve-dimensional integral (see for example
\cite{he1}). With the help of (\ref{poloi1}) we derived an
alternative representation of the purity by an infinite sum of
eight-dimensional integrals
\begin{eqnarray} \mbox{Tr}[\rho^2]=(2\pi)^4\sum_{l=0}^{\infty}\int_{0}^{\infty}...\int_{0}^{\infty} \int_{0}^{\pi}... \int_{0}^{\pi}(r r^{'}r_{1}r_{2})^2\nonumber\\
\psi(r,r_{1},\cos\theta)\psi(r^{'},r_{1},\cos\theta^{'})\psi(r,r_{2},\cos\theta^{''})\psi(r^{'},r_{2},\cos\theta^{'''})\nonumber\\P_{l}(\cos\theta)
P_{l}(\cos\theta^{'}) P_{l}(\cos\theta^{''})
P_{l}(\cos\theta^{'''})\sin\theta\sin\theta^{'}\sin\theta^{''}\nonumber\\\sin\theta^{'''}dr_{1}dr_{2}drdr^{'}d\theta
d\theta^{'}d\theta^{''}d\theta^{'''},\label{sukq}\end{eqnarray}
which may be useful when dealing with spherically
symmetric two-particle systems. If most of the
electrons' correlation is captured by the partial components with
low $l$, the sum (\ref{sukq}) may be more effective
than the mentioned
twelve-dimensional integral to calculate the purity.


The calculation of the vN entropy requires detemination of the
Schmidt coefficients. For singlet S states, we have
 in terms of the  occupation numbers
 $S=-\sum_{nl} (2l+1) \lambda_{nl} \log_{2} \lambda_{nl}$,
$L=1-\sum_{nl} (2l+1) \lambda_{nl} ^2$ \cite{9}. To determine the coefficients $\lambda_{nl}= ({4\pi k_{nl}\over 2l+1})^2$
we solved Eq. (\ref{poloi1})
through a discretization technique. A set of approximations to the $n_{m}+1$ coefficients $k_{nl}$ can
be thus obtained by diagonalizing  the matrix $[M_{ij}^{(l)}]$,
$M_{ij}^{(l)}=\bigtriangleup r f_{l}(\bigtriangleup r
i,\bigtriangleup r j)$, $\bigtriangleup r ={R/ n_{m}}$,
$i,j=0,...,n_{m}$.\cite{9}, where   $R$ should be chosen as large as
the side of a square in which the functions $f_{l}(r_{1},r_{2})$ are
mainly confined. Having the coefficients $k_{nl}$ determined in that
way for $l$ up to $l_{m}$, we obtain the approximate entropies $S=-\sum_{n=0}^{n_{m}}\sum_{l=0}^{l_{m}} (2l+1)
\lambda_{nl} \log_{2} \lambda_{nl}$ and
$L=1-\sum_{n=0}^{n_{m}}\sum_{l=0}^{l_{m}} (2l+1) \lambda_{nl}^2$.
In order to obtain stable numerical values, the calculations have to be repeated
for larger and larger  $R$ and $n_{m}$, $l_{m}$ until the results
converge to the desired accuracy.

\section{ Numerical results}\label{res}

In our  ground-state calculations  we employ the Hylleraas
variational wave function
  \be \psi(r_{1},r_{2},\cos\theta)=  \sum_{nmp}
c_{nmp}e^{-\mu s }s^{n}t^{m}u^{p},\label{funm}\ee with $0\leq
n+m+p\leq \omega$ ($m$-even), where   $s=r_{1}+r_{2}, t=r_{1}-r_{2},
u=r_{12}=|\vec{r}_{2}-\vec{r}_{1}|=(r_{1}^2+r_{2}^2-2 r_{1} r_{2}
\cos \theta)^{{1\over 2}}$ and $\mu$ is a non-linear variational
parameter.


The ground state energy $E$ and the
corresponding linear parameters $c_{ nmp}$ are determined by the solution of a
generalized eigenvalue problem \be
\sum_{nmp}(H_{n^{'}m^{'}p^{'},nmp}-E
S_{n^{'}m^{'}p^{'},nmp})c_{nmp}=0\ee where
$S_{n^{'}m^{'}p^{'},nmp}=\langle n^{'}m^{'}p^{'}|nmp\rangle$ and
$H_{n^{'}m^{'}p^{'},nmp}=\langle n^{'}m^{'}p^{'}|H|nmp\rangle$,
whereas, the non-linear parameter ƒ$\mu$ is iteratively optimized
so as  to minimize the approximate energy $\partial E^{(\omega)}
/\partial \mu=0$.

For  demonstration purpose, the $\omega$-order ground state energies
obtained as described above are shown in Table \ref{tab:table5pp},
where the underlines represent the digits that agree with the very
accurate results of Nakashima and Nakatsuji \cite{com}.

\begin{table}[h]
\begin{center}
\begin{tabular}{llllll}
\hline
$\omega$ & $Z=1$&$Z=2$ &$Z=3$&$Z=4$& $Z=5$ \\
\hline
$ 6$ & $-\underline{0.5277}432488 $ &$-\underline{2.90372}3702 $ &$-\underline{7.27991}2718    $&$-\underline{13.65556}549$& $-\underline{22.03097}079$ \\
$ 8$ & $-\underline{0.52775}00643 $ & $-\underline{2.9037243}05$&$-\underline{7.279913}342$&$-\underline{13.655566}16$& $-\underline{22.0309715}0$ \\
$ 10$ & $-\underline{0.52775}08656 $ &$-\underline{2.9037243}66$ &$ -\underline{7.2799134}02   $&$$& $$ \\
$ 12$ & $-\underline{0.52775}09860 $ &$-\underline{2.90372437}5$ &$    $&$$& $$ \\
$ 14$ & $-\underline{0.5277510}091 $ & &$    $&$$& $$ \\

\end{tabular}
\caption{\label{tab:table5pp}  Ground state energies determined
variationally as discussed in the text.}
\end{center}
\vspace{-0.6cm}
\end{table}

Calculating entanglement entropies, we first determine the Schmidt
coefficients by solving Eq. (\ref{integr}) numerically. We found
that for the Hylleras expansion, the integrals (\ref{poloi1}) can be
carried out analytically: Substitution of an explicit representation
$P_{l}(\cos\theta)=2^l \sum_{k=0}^{l} (\cos \theta)^{k} {l \choose
k}{{l+k-1\over 2} \choose l}$, and the Hylleras expansion
(\ref{funm}) expressed in $r_{1}, r_{2}$ and $\theta$ into
(\ref{poloi1}) yields
\begin{eqnarray}
f_{l}(r_{1},r_{2})=C {2^{l-1}(2l+1)}  r_{1} r_{2}
e^{-\mu (r_{1}+r_{2})}\sum_{k=0}^{l}\sum_{nmp}c_{nmp}{l \choose
k}{{l+k-1\over 2} \choose
l}(r_{2}-r_{1})^{m}(r_{1}+r_{2})^n I(k,p)\label{plp4}
\end{eqnarray}
where $C$ is the normalization constant and  $I(k,p)$ are given by
the following integrals \begin{eqnarray}  I(k,p)=
\int_{0}^{\pi}\sin\theta (\cos \theta)^{k}(r_{1}^2+r_{2}^2-2 r_{1}
r_{2} \cos \theta)^{{p\over 2}} d\theta ={1\over
r_{1}^2+r_{2}^2}\Gamma(1+k) \nonumber\\ ({(-1)^k
(r_{1}+r_{2})^{2+p}\over\Gamma(2+k)}{}_2F_{1}(1,2+k+{p\over 2},2+k,-
{2 r_{1}r_{2}\over r_{1}^2+r_{2}^2})
+{(r_{1}-r_{2})^{2+p}\over\Gamma(2+k)}{}_{2}F_{1}(1,2+k+{p\over
2},2+k, {2 r_{1}r_{2}\over
r_{1}^2+r_{2}^2})).\label{plkop}\end{eqnarray} 
In some cases it is computationally less
demanding to treat Eqs. (\ref{poloi1}) numerically for
discretized values of $r_{1}$ and $r_{2}$, especially when a large
number of terms is included in the Hylleras expansion. The above
analytical expressions of Eqs. (\ref{poloi1}) are however
useful when testing  the accuracies of  numerical integrations.

\begin{table}[h]
\begin{center}
\begin{tabular}{lllll}
\hline
$$ & $R=7$ &$R=9$&$R=10$ \\
\hline
$\omega=6  $ & $0.0159173$ & $0.0159162$&$0.0159162$ \\
$\omega=10  $ & $0.0159172$ & $0.0159157$&$0.0159157$\\
$\omega=14  $ & $0.0159172$ & $0.0159157$&$0.0159157$ \\

\hline
\end{tabular}
\caption{\label{tab:table1}The stable numerical results for the linear entropy $L$
obtained at different $R$ with different expansion lengths
$\omega=6, 10, 14$ corresponding to number of terms $50,  161, 372$,
respectively. }
\end{center}
\vspace{-0.6cm}
\end{table}

\begin{table}[h]
\begin{center}
\begin{tabular}{lllll}
\hline
$$ & $n_{m}=300$ &$n_{m}=600$&$n_{m}=1200$ \\
\hline
$ l_{m}=0$ & $0.0159242 $ & $0.0159207 $&$0.0159205 $ \\

$ l_{m}=1$ & $0.0159194$ & $0.0159159 $&$0.0159157$ \\
$ l_{m}=2$ & $0.0159194$ & $0.0159159 $&$0.0159157$ \\

\hline
\end{tabular}
\caption{\label{tab:table2} The linear entropy $L$ computed at
$R=10$ with an expansion given by a 372-term wavefunction (
$\omega=14$) as a function of $l_{m}$, for $n_{m}=300,600,1200$
corresponding  to $\bigtriangleup r=30^{-1},60^{-1}, 120^{-1}$,
respectively. }
\end{center}
\vspace{-0.6cm}
\end{table}

\begin{table}[h]
\begin{center}
\begin{tabular}{lllll}
\hline
$$ & $n_{m}=300$ &$n_{m}=600$&$n_{m}=1200$ \\
\hline
$ l_{m}=0$ & $0.0428655$ & $0.0428631$&$0.0428630$ \\
$ l_{m}=1$ & $0.0814955$ & $0.0814931$&$0.0814930 $ \\
$ l_{m}=2$ & $0.0842412$ & $0.0842388$&$0.0842387$ \\
$ l_{m}=3$ & $0.0847083$ & $0.0847058$&$0.0847057$ \\
$ l_{m}=4$ & $0.0848295$ & $0.0848271$&$0.0848269$ \\
$ l_{m}=5$ & $0.0848702$ & $0.0848678$&$0.0848676$ \\
$ l_{m}=10$ & $0.0849006$ & $0.0848982$&$0.0848980$ \\
$ l_{m}=14$ & $0.0849022$  & $0.0848997$&$0.0848996$ \\
$ l_{m}=18$ & $0.0849025$ & $0.0849001$&$0.0848999$ \\
$ l_{m}=20$ & $0.0849025$ & $0.0849001$&$0.0848999$ \\

\hline
\end{tabular}
\caption{\label{tab:table3}Same as in Table \ref{tab:table2}, but
for the von Neumann entropy $S$. }
\end{center}
\vspace{-0.6cm}
\end{table}
 In order to gain
insight into the effectiveness of the method described in previous
section, we first determine the  occupation numbers of the ground
state helium atom and assess their accuracy by comparing the linear
entropy  with the data available in literature. 
 Our numerical values obtained for $L=1-\sum_{nl} (2l+1) \lambda_{nl}^2$ at different $R$ with different
expansion lengths $\omega$ are listed  in table \ref{tab:table1}.
The numerical stability was achieved by increasing $n_{m}$ and $l _{m}$ until the results stay fixed to the quoted accuracy.
 It can be seen that already at $R=9$ and $\omega=10$ the results start to
match with the benchmark value for the linear entropy $0.0159156 \pm
0.000001$ established with relative small estimated uncertainty in
Ref. \cite{he4},
 which
 proves the effectiveness of the method we are using here for
determining the occupation numbers. Tables \ref{tab:table2} and
\ref{tab:table3} show how the values of the linear and von Neumann entropies, respectively, converge
 as the cut-offs $l_{m}$ and $n_{m}$ are increased. Performing calculations at larger $R$ and $\omega$,
we have verified that the value of the vN entropy $0.0848999$
faithfully reproduces the true value with to at least $7$
significant digits. It is worth stressing at this point that the convergence with increasing $l_{m}$  appears monotonic (from above for the linear entropy and from below for the vN entropy).

In Table \ref{tab:table4} our results for the entropies of the helium atom are  compared with those obtained by other workers  in different ways. For example, in
Refs.
 \cite{he3} and \cite{he5} the authors   calculated the vN and linear  entropies  using
configuration interaction method with  basis wave functions
constructed by Slater-type orbitals  (STO) and by B-spline basis
with one-electron momentum states  up to $l=3$ and up to $l=5$,
respectively. 
From the comparison, we conclude that our value for the vN
entropy is the best so far determined for the helium
atom.

We also computed the entanglement entropies for the ground states of
the two-electron atoms with different values of $Z$. Our results for
the linear entropy and the vN entropy are listed in the table
\ref{tab:table5}, where a comparison with the literature
\cite{he2,he4} is also made. It is worth stressing  that in each
case considered here, the stability of the results up to at least
six decimal places  was achieved  already at $l_{m}=1$, similarly as
for the helium atom. Our value of the linear entropy of the hydrogen
negative ion $(Z=1)$ coincides with the recently obtained value
$0.106153$ of Ref. \cite{he4}. In all the remaining cases, our
values are more accurate being slightly lower  than the results of
the recent calculations \cite{he2}. The only accurate value reported
in the literature is that for the helium atom \cite{he5} which
compares well with our result. The vN entropy for other values of
$Z$ was calculated only in Ref. \cite{Hofer}, where the convergence
of correlated Gaussian basis sets has been tested. However, the
results for helium-like ions reported in supplementary material to
this work differ widely depending on the type of Gaussian basis
used. Despite of using large basis sets, the results for the linear
and vN entropies of helium-like ions obtained in Ref. \cite{Hofer}
are of low accuracy and were presented only graphically in the
article. We would like to stress that using the Hylleraas basis set
(\ref{funm}) provides much better convergence properties, which
enabled us to determine the vN entropy to 6 digits accuracy.

\begin{table}[h]

\begin{center}
\begin{tabular}{cccc}
\hline

& &$L$ & $S$ \\
\hline

& This work & 0.0159157& 0.0848999\\

& Dehesa  et al \cite{he1}  & $0.015914\pm0.000044$ & \\

& Benetti  et al  \cite{he3} & 0.01606 & 0.0785\\

& Y. C. Lin et al \cite{he2}   & $0.015943\pm0.00004$ & \\
& C. H. Lin et al \cite{he4}  & $0.0159156 \pm 0.000001$ & \\
& Y.C Lin et al \cite{he5}  & $0.015943$ & 0.085022\\
&Huang et al \cite{huang} &$$& 0.0675\\

\hline

\end{tabular}
\caption{\label{tab:table4}Comparison of the vN and linear entropies calculated for the helium atom ground state
with the results published in literature.}
\end{center}
\end{table}

\begin{table}[h]
\begin{center}
\begin{tabular}{llllll}
\hline
$$ & $Z=1$&$Z=2$ &$Z=3$&$Z=4$& $Z=5$ \\
\hline\\
$ L$ & $0.106153$ &0.0159157 &$ 0.006539   $&$0.003558$& $0.002235$ \\
\cite{he2} & & $0.015943 $&$0.006549$&$0.003562$& $0.002237$ \\
\cite{he4}& 0.106153& $0.0159156 $&$$&$$& $$ \\
\hline
$ S$ & $0.380012$&0.0848999 &$0.039496$&$0.023146$&$0.015324$ \\
\cite{he5}& & $0.085022$&$$&$$& $$ \\

\hline
\end{tabular}
\caption{\label{tab:table5} Linear entropy (L) and the vN entropy
(S) calculated for the ground state of helium-like ions compared
with the best literature results.}
\end{center}
\vspace{-0.6cm}
\end{table}

The accurate results allow us to study the relation between the
linear entropy and the vN entropy of the RDM for the helium-like
ions as a function of $Z$. This is an important issue since the
linear entropy is frequently used to measure entanglement in the
system. The linear entropy  is much easier to calculate than the vN
entropy since it is directly calculable from the integral
representation and does not require diagonalization of the RDM.

Comparing the dependence on $Z$ of the linear and vN entropies, we
noted that from $Z=2$ to $Z=5$ they are almost linearly related. {
This is demonstrated in Fig.\ref{fig:odog1}, where the vN entropy is
shown together with the rescaled linear entropy $6.856 L$, where the
factor $6.856$ is obtained as the proportionality constant between
$S$ and $L$ at $Z=5$. The departure from the linear realtionship
occurs in the vicinity of $Z=1$. This may be caused by the proximity
to the critical point $Z_{c}$ below which there are no bound states
in the system. It has been shown in Ref. \cite{cri} that the
ionization point at which the helium-like system has a bound state
with zero binding energy is at $Z_{c}\approx 0.911$. Our calculation
show that in the vicinity of the critical point, where the system is
highly correlated, the behavior of the linear and vN entropies is
different.

Our results may be compared with the discussion of entanglement in the spherical helium-like model in Ref. \cite{osenda}.
The spherical model is an approximation to the atom  obtained by replacing the Coulombic repulsion between the electrons
by its spherical average. The approximate model has been shown to exhibit a similar near-threshold behavior as the two-electron atom \cite{serra}.
 In Ref. \cite{osenda}, the scaling properties of the von Neumann entropy have been studied for the ground state
 of the spherical helium-like model and its singular behavior was demonstrated at $Z_{c}^{sph}\approx 0.949$. This value appears really close to the critical point of the helium-like atom ($Z_{c}\approx 0.911$).  

\begin{figure}[h]
\begin{center}
\includegraphics[width=0.50\textwidth]{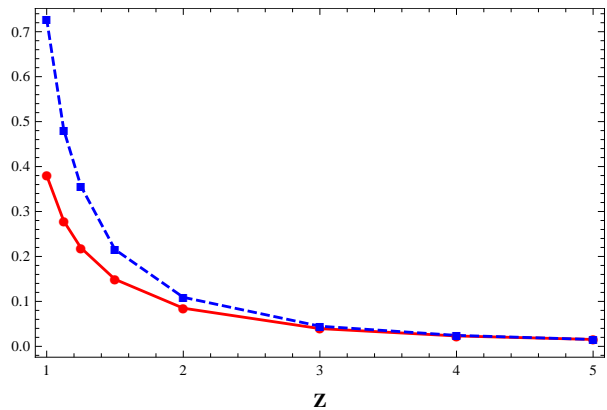}
\end{center}
\caption{Comparison of the vN entropy (full line) and the rescaled linear entropy
$6.856L$ (dashed line) as functions of $Z$. } \label{fig:odog1}
\end{figure}

\section{Conclusions}\label{summ}
In conclusion, we have performed accurate calculations of the linear
and vN entropy of the ground states of the helium  atom and
helium-like ions, basing on the Schmidt decomposition of the
two-particle spatial wavefunctions. The accurate wvefunctions were
obtained, employing expansions in terms of original Hylleraas-type
basis functions. Using a discretization technique, we determined the
natural occupation numbers $\lambda_{nl}$ up to very large $l$ and
$n$ for a series of values of the nuclear charge from $Z=1$ to $Z=5$,
which enabled high-precision determination of the corresponding
entropies. In particular, the vN entropies of the helium-like ions
have been calculated for the first time and that of the helium atom
has been determined with much better accuracy than earlier
calculations. Furthermore, our results revealed that relationship
between the vN and linear entropies is almost linear for $Z\geq2$.
However with $Z$ decreasing to the critical value, the increase of
the vN entropy gets much faster than that of the linar entropy. This
may give a warning that using the linar entropy instead of the vN
entropy to measure entanglement not always is justified.

\bibliography{aipsamp}\nonumber

\begin{thebibliography}{99}
\bibitem{2} Ya\~{n}ez, R., A. Plastino, and J. Dehesa: Quantum entanglement in a soluble two-electron model atom, Eur. Phys. J. D 56
141 (2010)
\bibitem{3} Majtey, A., A. Plastino,  and J. Dehesa:The relationship between entanglement, energy and level degeneracy in two-electron systems  J. Phys. A: Math. Theor. 45
 115309 (2012)
\bibitem{3.1} Bouvrie, P. A.  et al.:Quantum entanglement in exactly soluble atomic models: the Moshinsky model with three electrons, and with two electrons in a uniform magnetic field, Eur. Phys. J. D 66
15 (2012)
\bibitem{4.1} Ko\'{s}cik P and Okopi\'{n}ska:Correlation effects in the Moshinsky model, Few-Body Syst 54
1637 (2013)

\bibitem{4} Manzano, D., et al.:Quantum entanglement in two-electron atomic models,  J. Phys. A Math. Theor. 43
275301 (2010)


\bibitem{5} Coe, J., A. Sudbery, and I. D'Amico:Entanglement and density-functional theory: Testing approximations on Hooke's atom, Phys. Rev. B 77
 205122 (2008)

\bibitem{7} Ko\'{s}cik, P.:Two-electron entanglement in a two-dimensional isotropic harmonic trap: Radial correlation effects in the low density limit, Phys. Lett. A 375
458 (2011)
\bibitem{8} Ko\'{s}cik, P., and H. Hassanabadi:Entanglement in Hooke's Law Atoms: an Effect of the Dimensionality of the Space, Few-Body Systems  52
189-192 (2012)
\bibitem{9} P. Ko\'{s}cik:Entanglement in S states of two-electron quantum dots with Coulomb impurities at the center , Phys, Lett. A 377
2393 (2013)
\bibitem{9.1} Ko\'{s}cik P and Okopi\'{n}ska A:Two-electron entanglement in elliptically deformed quantum dots,   Phys. Lett. A 374
3841 (2010)

\bibitem{10} Nazmitdinov, R. et al.: Shape transitions in excited states of two-electron quantum dots in a magnetic field, J. Phys. B: At. Mol. Opt. Phys. 45
 205503 (2012)

\bibitem{11} Acosta Coden, D. et al.:Impurity effects in two-electron coupled quantum dots: entanglement modulation, J. Phys. B: At. Mol. Opt. Phys. 46
065501 (2013)

\bibitem{vn}R. Pa\v{s}kauskas,L. You, Quantum correlations in
two-boson wave functions,  Phys. Rev. A 64, 042310 (2001)


\bibitem{he1}J. S. Dehesa, et al.:Quantum entanglement in helium, J. Phys. B: At. Mol. Opt.
Phys. 45, 015504 (2012).

\bibitem{he3}G. Benenti, S. Siccardi, G. Strini:Entanglement in helium, Eur. Phys. J. D
67–83 (2013)

\bibitem{he2}Y.C Lin, C.Y Lin, and Y.K. Ho:Spatial entanglement in two-electron atomic systems, Phys. Rev. A 87, 022316 (2013)

\bibitem{he4} C.H. Lin, C.Y. Lin, and Y. K. Ho: Quantification of Linear Entropy for Quantum Entanglement in He, H- and Ps- Ions Using Highly-Correlated Hylleraas Functions,Few-Body
Systems  54,  2147 (2013)

\bibitem{he5}Y.C Lin, Y. K. Ho:,Quantum entanglement for two electrons in the excited states of helium-like systems
arXiv:1307.5532 (2013)
\bibitem{Wang} Jia Wang, C. K. Law, and M.-C. Chu: s-wave quantum entanglement in a harmonic trap, Phys. Rev. A 72, 022346 (2005)
\bibitem{sh}E.R. Davidson:Natural expansions of Exact Wavefunctions. III. The Helium-Atom Ground State , The Journal of Chemical Physics, 39,875
(1964)

\bibitem{pla} Plastino, A., D. Manzano, and J. S. Dehesa:Separability criteria and entanglement measures for pure states of N identical fermions, Europhys. Lett. 86
20005 (2009)

\bibitem{huang} Z. Huang, H. Wang and S. Kais:Entanglement and electron correlation in quantum chemistry calculations, Journal of Modern Optics 53, 2543 (2006).

\bibitem{Hofer} Thomas S.Hofer:, On the basis set convergence of electron–electron entanglement measures: helium-like systems, Front. Chem. 1:24. (2013)

\bibitem{com} H. Nakashima and H. Nakatsuji,Solving the Schr\"{o}dinger equation for helium atom and its isoelectronic
ions with the free iterative complement interaction (ICI) method, J.
Chem. Phys. 127, 224104 (2007)


\bibitem{cri}J.D. Baker, E. Freund, R.N. Hill, J.D. Morgan::Radius of convergence and analytic behavior of the  expansion, Phys. Rev. A 41
 1247 (1990)
\bibitem{osenda} O. Osenda and P. Serra:Scaling of the von Neumann entropy in a two-electron system near the ionization threshold, Phys. Rev. A 75
042331 (2007)
\bibitem{serra}P. Serra: Comment on "Towards a differential equation for the nonrelativistic ground-state electron density of the He-like sequence of atomic ions", Phys. Rev. A 74
016501 (2006)












\end{thebibliography}

\end{document}